\documentclass[reprint,superscriptaddress,twocolumn,amsmath,amssymb,aps,prl]{revtex4-2}
\usepackage{graphicx}
\usepackage{amsmath}
\usepackage{amssymb}
\usepackage{braket}
\usepackage{dcolumn}
\usepackage{upgreek}
\usepackage{tikz}
\usepackage[colorlinks=true,linkcolor=blue,citecolor=blue,urlcolor=blue]{hyperref}
\usepackage{natbib} 
\usepackage[normalem]{ulem}

\newcommand{\uK}{\rm\upmu K}
\newcommand{\mK}{\rm mK}
\newcommand{\um}{\rm\upmu m}

\newcommand{\MHz}{\rm MHz}
\newcommand{\mW}{\rm mW}
\newcommand{\W}{\rm W}
\newcommand{\degr}{^\circ}

\begin{document}

\title{Observation of Near-Critical Kibble-Zurek Scaling in Rydberg Atom Arrays}
\author{Tao Zhang}
\thanks{These authors contributed equally to this work.}
\affiliation{Department of Physics and State Key Laboratory of Low Dimensional Quantum Physics, Tsinghua University, 100084, Beijing, China.}
\author{Hanteng Wang}
\thanks{These authors contributed equally to this work.}
\affiliation{Institute for Advanced Study, Tsinghua University, 100084, Beijing, China.}
\author{Wenjun Zhang}
\thanks{These authors contributed equally to this work.}
\author{Yuqing Wang}
\thanks{These authors contributed equally to this work.}
\author{Angrui Du}
\author{Ziqi~Li}
\author{Yujia Wu}
\affiliation{Department of Physics and State Key Laboratory of Low Dimensional Quantum Physics, Tsinghua University, 100084, Beijing, China.}
\author{Chengshu Li}
\affiliation{Institute for Advanced Study, Tsinghua University, 100084, Beijing, China.}
\author{Jiazhong Hu}
\email{hujiazhong01@ultracold.cn}
\affiliation{Beijing Academy of Quantum Information Sciences, 100193, Beijing, China.}
\author{Hui Zhai}
\email{hzhai@tsinghua.edu.cn}
\affiliation{Institute for Advanced Study, Tsinghua University, 100084, Beijing, China.}
\author{Wenlan Chen}
\email{cwlaser@mail.tsinghua.edu.cn}
\affiliation{Department of Physics and State Key Laboratory of Low Dimensional Quantum Physics, Tsinghua University, 100084, Beijing, China.}
\affiliation{Frontier Science Center for Quantum Information and Collaborative Innovation Center of Quantum Matter, 100084, Beijing, China.}
\date{\today}

\begin{abstract}
The Kibble-Zurek scaling reveals the universal dynamics when a system is linearly ramped across a symmetry-breaking phase transition. However, in reality, inevitable finite-size effects or symmetry-breaking perturbations can often smear out the critical point and render the phase transition into a smooth crossover. In this letter, we show experimentally that the precise Kibble-Zurek scaling can be retained in the near-critical crossover regime, not necessarily crossing the critical point strictly. The key ingredient to achieving this near-critical Kibble-Zurek scaling is that the system size and the symmetry-breaking field must be appropriately scaled following the variation of ramping speeds. The experiment is performed in a reconfigurable Rydberg atom array platform, where the Rydberg blockade effect induces a $Z_2$ symmetry-breaking transition. The atom array platform enables precise control of the system size and the zigzag geometry as a symmetry-breaking field. Therefore, we can demonstrate notable differences in the precision of the Kibble-Zurek scaling with or without properly scaling the system size and the zigzag geometry. Our results strengthen the Kibble-Zurek scaling as an increasingly valuable tool for investigating phase transition in quantum simulation platforms.

\end{abstract}

\maketitle

The most remarkable feature of phase transitions is that systems with very different microscopic physics can exhibit universal scaling laws as they approach the critical point. The nonequilibrium physics of driving a system across a symmetry-breaking phase transition was first discussed by Kibble in the context of cosmological phase transitions~\cite{ kibble1976topology} and by Zurek in the thermal phase transitions of condensed matter systems~\cite{zurek1985cosmological}. They reveal a highly nontrivial fact that although the universal scaling exponents are introduced through equilibrium properties, they can also emerge in the nonequilibrium ramping dynamics. Precisely, the correlation lengths measured after ramping versus the ramping speeds were predicted to exhibit a universal scaling law, now known as the celebrated Kibble-Zurek (KZ) scaling~\cite{zurek1996cosmological,laguna1997density}. Later, KZ scaling was also generalized to quantum phase transitions~\cite{zurek2005dynamics, dziarmaga2005dynamics,polkovnikov2005universal,polkovnikov2011colloquium,del2014universality,rossini2021coherent}.

KZ scaling has attracted considerable experimental interest over decades, ranging from classical to quantum systems~\cite{chuang1991cosmology,hendry1994generation,ruutu1996vortex, ulm2013observation, pyka2013topological, navon2015critical, clark2016universal,keesling2019quantum,ebadi2021quantum,lee2024universal,Huang2021KZ,Zheng2023KZ,manovitz2025quantum,king2022coherent,king2023quantum,li2023probing}. Most recent experimental tests of KZ scaling have been performed on various quantum simulation platforms, including ultracold atoms~\cite{navon2015critical,clark2016universal, keesling2019quantum,ebadi2021quantum,lee2024universal,Huang2021KZ,Zheng2023KZ,manovitz2025quantum}, superconducting qubits~\cite{king2022coherent,king2023quantum}, and trapped ions~\cite{ulm2013observation, pyka2013topological, li2023probing}, as the parameters of these platforms can be precisely controlled. In condensed matter systems, the critical exponents of phase transitions are often measured by studying how the static properties, such as susceptibility and specific heat, diverge as one approaches critical points. KZ scaling has become an increasingly important dynamic tool for measuring critical exponents in quantum simulation platforms, complementary to static property measurements. 

\begin{figure*}
    \begin{center}  
    \includegraphics[width=1.4\columnwidth]{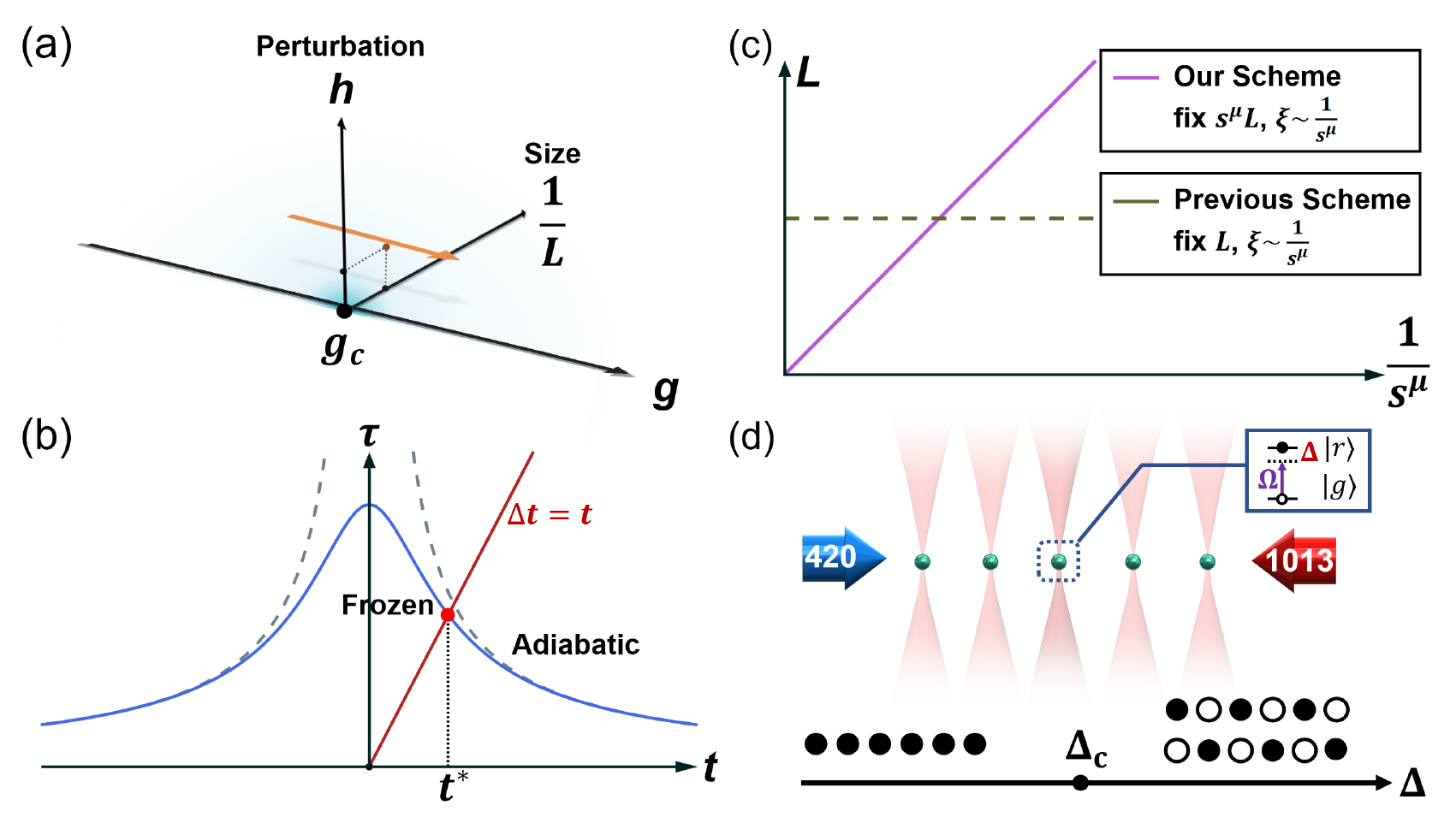}
    \caption{The principle of our experiment. (a) The key point of this work: a strict quantum critical point characterized by divergent response time can be smeared out by finite size or destroyed by symmetry-breaking perturbations, resulting in a crossover regime near the critical point. Here, we consider ramping in the nearby crossover regime instead of strictly crossing the quantum critical point. A precise KZ scaling can still be obtained, provided that we scale the system size and perturbation strength properly following the change of ramping speeds $s$. (b)   Illustration of the physical mechanism for KZ scaling. The characteristic time scale $\tau$ diverges at the critical point only for an infinitely large system (dashed line) and follows a universal yet regular functional form for any finite-size system (solid line). The red line $\Delta t$ intersects $\tau$ at $t^*$, dividing the dynamics into the frozen and adiabatic regimes, respectively. (c) Comparison between our experiment and previous experiments on KZ scaling. Our experiment verifies $\xi \sim 1/s^\mu$ scaling along the purple solid line with $s^\mu L$ fixed, while earlier experiments study the scaling along the dashed line with a fixed system size $L$. (d) Illustration of the experimental system of Rydberg atom arrays and two phases with a $Z_2$ phase transition driven by tuning $\Delta$.    } 
    \label{illustration}
        \end{center}
\end{figure*}

However, the finite-size effect can eliminate the singularity at critical points, which is often pronounced in these quantum simulation platforms. Additionally, symmetry-breaking perturbations frequently occur, explicitly destroying the symmetry at the Hamiltonian level. As illustrated in Fig.~\ref{illustration}(a), both the finite-size effect and the symmetry-breaking perturbations can drive the system away from the critical point, entering into a surrounding crossover regime. In reality, the ramping trajectory often passes through the near-critical crossover regime rather than strictly crossing the critical point. Therefore, a key issue is whether these effects inevitably lead to deviations from KZ scaling or whether a protocol exists to eliminate these derivations and restore precise KZ scaling even in the near-critical crossover regime. So far, only a few theoretical works have paid attention to the finite size effect of the ramping dynamics~\cite{Sandvik2011,rossini2021coherent,de2023out}, and the experimental investigation of this issue is still lacking.    

\textit{Summary of Theoretical Protocol.} Suppose the system undergoes a second-order phase transition by tuning a dimensionless parameter $g$, entering a symmetry-breaking phase at $g>g_\text{c}$, and we consider a linear ramp that $\delta g\equiv g-g_\text{c}$ changes as $st$ with a constant ramping speed $s$. 
The original KZ scaling states that the correlation length $\xi$ measured after ramping scales with $s$ as $1/s^\mu$, and it originates from the so-called critical slowing down, namely, the relaxing time $\tau$ diverges as $\tau\sim 1/\delta g^{z\nu}$ as approaching the critical point~\cite{zurek1985cosmological,zurek1996cosmological,laguna1997density,del2014universality}. Here, both $z$ and $\nu$ are universal critical exponents and $\mu=\nu/(1+z\nu)$. 

However, in a finite-size system, the relaxation time does not diverge even at $g_\text{c}$, as shown in Fig.~\ref{illustration}(b). Nevertheless, $\tau$ depends on system size $L$ through a universal function form as $\tau=L^{z} [\mathcal{F}(\delta g L^{1/\nu})]^z$. As we will derive in the End Matter, although the explicit function form of $\mathcal{F}(x)$ is not known, this functional dependence is sufficient to conclude that 
\begin{equation}
\xi\sim \frac{1}{s^\mu}f(s^\mu L).
\end{equation} 
Although the explicit function form of $f(x)$ is also unknown, it is important that this function only depends on $s^\mu L$. Therefore, the protocol is that when we vary the ramping speed to explore the KZ scaling, the system size must be adjusted accordingly to keep $s^\mu L$ constant, as illustrated by the solid line in Fig.~\ref{illustration}(c). This approach ensures a precise $\xi\sim 1/s^\mu$ unaffected by finite-size effects. This contrasts with the previous scheme, as shown by the dashed line in Fig.~\ref{illustration}(c). Previous experiments always measured the KZ scaling with a fixed system size. It could lead to deviation from the expected critical exponents, especially when $L$ is not large enough or when $s$ is not large enough.

\begin{figure}
    \begin{center}
    \includegraphics[width=0.9\columnwidth]{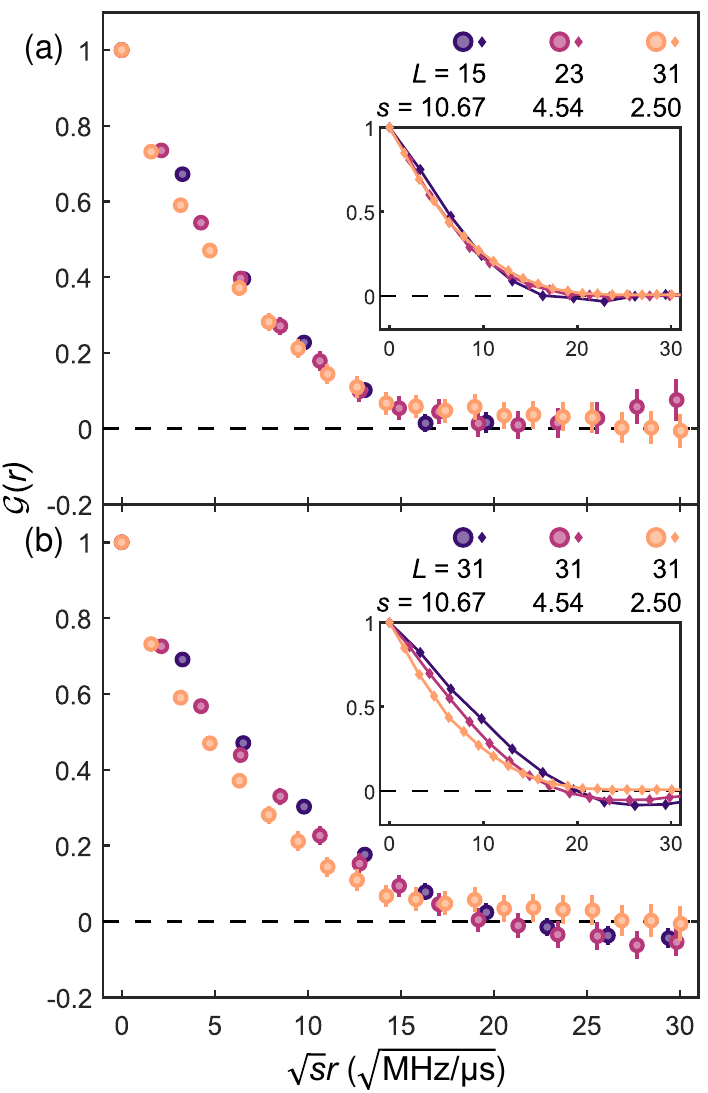}
    \caption{Comparison between two schemes for measuring KZ scaling. The normalized two-point correlation function $\mathcal{G}(r)$ (see text for definition) measured after ramping is plotted as a function of $\sqrt{s}r$, for three different ramping speeds $s=10.67~\text{MHz}/\mathrm{\upmu s}$, $4.54~\text{MHz}/\mathrm{\upmu s}$ and $2.50~\text{MHz}/\mathrm{\upmu s}$. (a) Following our scheme, the system size varies with speed change, with $L=31$, $23$ and $15$, respectively. (b) Following previous schemes, the system size is fixed at $L=31$. The insets show the numerical simulation results~(see \cite{supplementary}) under the same parameters.     } 
    \label{comparison}
    \end{center}
\end{figure}

In addition to the finite size effect, a quantum critical point can also be smeared out by a perturbation $h$ that explicitly breaks the symmetry, as shown in Fig.~\ref{illustration}(a). The response time $\tau$ then also depends on $h$ as $\tau=L^z[\mathcal{F}(\delta g L^{1/\nu},hL^{1/\nu_h})]^z$, where $\nu_h$ is another critical exponent associated with the field $h$. Following the same strategy explained in the End Matter, it can be concluded that, in order for the correlation length to scale with the ramping speed $s$ as $1/s^\mu$, we must not only fix $s^\mu L$ but also fix $h L^{1/\nu_h}$. That is to say, we need to vary $h$ as $s^{\mu/\nu_h}$~\cite{zurek-bias}. 

\textit{Experimental Platform.} We implement our protocol in a reconfigurable atom array system using optical tweezers with controllable Rydberg interactions, as shown in Fig.~\ref{illustration}(d). This platform has become a playground for studying rich quantum many-body physics, especially non-equilibrium quantum dynamics~\cite{bernien2017probing,de2019observation, keesling2019quantum, ebadi2021quantum, altman2021quantum, kaufman2021quantum, bluvstein2021controlling, scholl2021quantum, young2022tweezer, zhao2024observation,fang2024probing,manovitz2025quantum}. KZ scaling has also been experimentally investigated in this platform~\cite{keesling2019quantum, ebadi2021quantum,manovitz2025quantum}.

In our experiment,  $^{87}$Rb atoms are initialized in the ground state hyperfine level $\ket{g}=\ket{5S_{1/2},F=2, m_F=+2}$ state by optical pumping. After switching off the tweezers, two counter-propagating lasers at $420$~nm and $1013$~nm are turned on to drive a coherent two-photon transition from $\ket{g}$ to the Rydberg state $\ket{r}=\ket{68S_{1/2}, m_J=+1/2}$~(see \cite{supplementary}). The resulting many-body dynamics is described by the following Hamiltonian, 
\begin{equation}
    \hat{H} = \frac{\Omega(t)}{2} \sum_i (\ket{r_i}\bra{g_i}+\text{h.c.}) - \Delta(t) \sum_i \hat{n}_i + \sum_{i<j}V_{ij} \hat{n}_i \hat{n}_j
\end{equation}
where $\hat{n}_i=\lvert r_i\rangle \langle r_i\lvert$ is the projection operator onto the Rydberg state, $V_{ij}$ is the $1/r^6$ van der Waals interaction between Rydberg states, and $\Omega$ and $\Delta$ are the two-photon Rabi frequency and detuning, respectively. 

We first create a one-dimensional defect-free array of single atoms. We turn on $\Omega$ to $\Omega=2$~MHz over $1~\mathrm{\upmu s}$ with an initial detuning $\Delta_\text{i} =-4$~MHz. With this value of Rabi frequency, the Rydberg blockade radius $R_\text{b}=1.5a$ for the spacing $a=5.3~\upmu\mathrm{ m}$ between the two nearest tweezers, preventing two Rydberg atoms setting at the neighboring sites. In this situation, the system experiences a phase transition at $\Delta_\text{c}/\Omega=1.1$ to a $Z_2$ symmetry-breaking phase at $\Delta>\Delta_c$, where the atoms arrange alternately between the ground and the Rydberg state, breaking the translational symmetry and doubling the unit cell, as shown in Fig.~\ref{illustration}(d). To study KZ scaling,  we linearly ramp $\Delta(t)$ with different ramping speeds $s$ to the final detuning $\Delta_\text{f} = 7.5$~MHz. Finally, we slowly turn off $\Omega$ over 1 $\mathrm{\upmu s}$ and measure the final state by turning back on the tweezers and performing the fluorescence imaging, which only detects atoms remaining in $\ket{g}$ after the entire process~(see \cite{supplementary}). 

Using the fluorescence snapshots obtained at the end of each experimental sequence, we calculate the Rydberg density-density correlation function as $G(r) = \langle \hat{n}_i \hat{n}_j \rangle - \langle \hat{n}_i \rangle \langle \hat{n}_j \rangle$ with $r=r_i-r_j$ fixed. To characterize the $Z_2$ symmetry breaking in this system, we introduce a normalized $\mathcal{G}(r)$ as $\mathcal{G}(r)=(-1)^r G(r)/G(0)$. Each $\mathcal{G}(r)$ is averaged over approximately $200-300$ snapshots and over $r_i$ for each snapshot~(see \cite{supplementary}). It is reasonable to assume that the correlation function around the critical point exhibits spatial scaling symmetry and $\mathcal{G}(r)$ depends only on $r/\xi$. Hence, if $\xi \sim s^\mu$, it implies that $\mathcal{G}(r)$ obtained with different ramping speeds should collapse into a single curve when they are plotted as a function of $s^\mu r$.

\textit{Verification of Finite-Size KZ Scaling.}  For $1+1$-dimensional Ising transition, we have $z=\nu=1$ and therefore $\mu=1/2$. Below, we first verify the finite-size KZ scaling in Fig. \ref{comparison} assuming $\mu=0.5$ a priori.

The experimental results of $\mathcal{G}(r)$ as a function of $\sqrt{s}r$ are presented in Fig.~\ref{comparison}. 
As illustrated in Fig.~\ref{comparison}(b), when $L$ is constant, the data deviates from $\sqrt{s}$ scaling as $s$ becomes small. Especially for $s=2.50~\text{MHz}/\mathrm{\upmu s}$, the ramping speed is already slower compared to the data reported in previous experiments~\cite{keesling2019quantum}, and we find that the correlation length is noticeably shorter than $\sqrt{s}$ scaling. This is consistent with what we expect of the finite-size effect, as explained in the End Matter. 
Nevertheless, for the same parameter of $L=31$ and $s=2.50~\text{MHz}/\mathrm{\upmu s}$, it is evident that when we vary the system size in accordance with the speed change to maintain $\sqrt{s}L$ constant, and even by including much smaller system sizes like $L=23$ and $L=15$,  Fig.~\ref{comparison}(a) shows nearly perfect data collapse,  significantly better than Fig.~\ref{comparison}(a) with a fixed $L$. 

\textit{A General Protocol to Extract $\mu$.} Then, we present a general method based on the finite-size scaling protocol to extract $\mu$, supposing $\mu$ is unknown.  
We measure the correlation function after ramping and obtain $\mathcal{G}_1(r)$ for $\{L_1,s_1\}$ and $\mathcal{G}_2(r)$ for $\{L_2, s_2\}$. These two datasets share the same value of $s^\mu L$ when $\mu$ is chosen as $\mu=\log(L_2/L_1)/\log(s_1/s_2)$. Then, we plot $\mathcal{G}_1$ and $\mathcal{G}_2$ as a function of $\tilde{r}=s^{\mu}r$ and use a function $\mathcal{D}$ defined below as a metric to characterize the distance between the two curves $\mathcal{G}_1(\tilde{r})$ and $\mathcal{G}_2(\tilde{r})$. This metric quantifies how well these two curves collapse into a single curve. We use this metric instead of fitting correlation length because we find that $\mathcal{G}(\tilde{r})$ does not strictly follow an exponential behavior, and extracting correlation length can lead to a certain degree of ambiguity. 

\begin{figure}
    \begin{center}  
    \includegraphics[width=0.9\columnwidth]{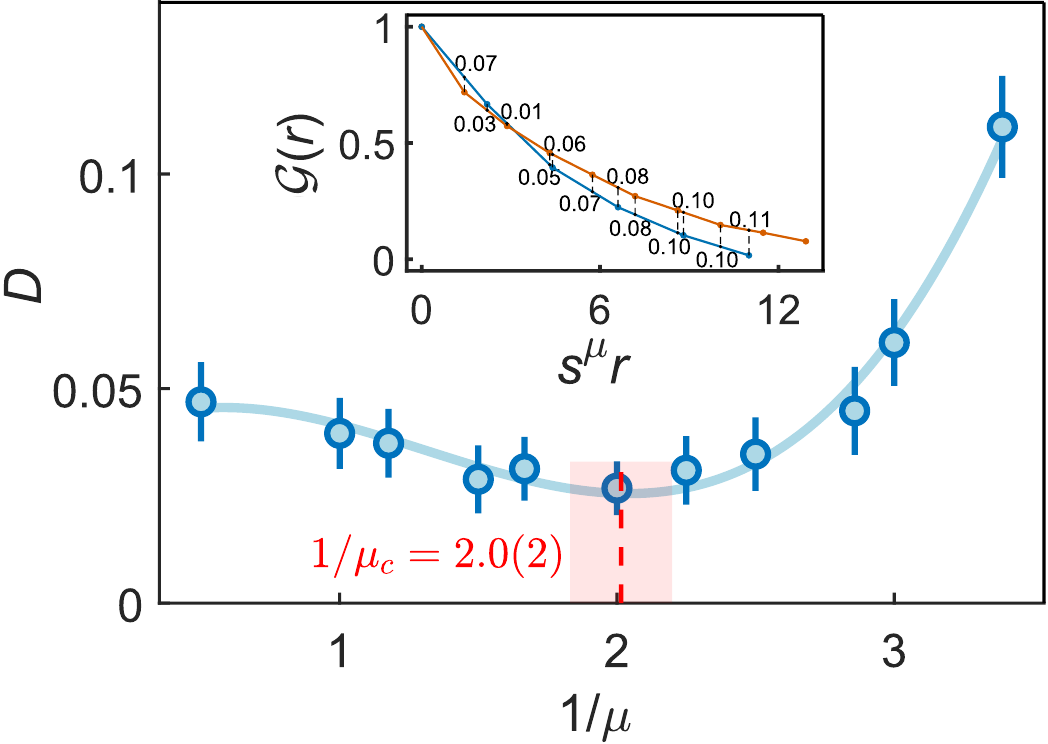}
    \caption{Extracting the critical exponent. The distance function $\mathcal{D}$, defined by Eq.~(\ref{distance}), measures the difference between two correlation functions rescaled by $s^\mu r$. Here, we set $L_1=15$, $s_1=10.67~\text{MHz}/\upmu \text{s}$, and $L_2=23$. We vary $s_2$ from $2.5~\text{MHz}/\upmu \text{s}$ to $8.62~\text{MHz}/\upmu \text{s}$ to adjust $\mu$. The blue line represents a typical cubic fit of $\mathcal{D}$, whose minimum determines the optimal value of $1/\mu_\text{c}$. The shaded red regime denotes the error bar of the optimal $1/\mu_\text{c}$. The inset illustrates an example of calculating $\mathcal{D}$ between two rescaled curves, with $s_2=2.96~\text{MHz}/\upmu \text{s}$ and $\mu$=1/3.} 
    \label{exponent}
    \end{center}
\end{figure}

The distance function $\mathcal{D}$ is introduced as 
\begin{equation}
\mathcal{D} = \frac{\sum_i |\mathcal{G}_1(\tilde{r}_i) - \mathcal{G}_2^{\text{lin}}(\tilde{r}_i)| + \sum_j |\mathcal{G}_2(\tilde{r}_j) - \mathcal{G}_1^{\text{lin}}(\tilde{r}_j)|}{N_{1} + N_{2}}. \label{distance}
\end{equation}
The function $\mathcal{G}_i^{\text{lin}}(\tilde{r})$ is the linear interpolation of $\mathcal{G}_i(\tilde{r})$, and $N_{1} + N_{2}$ in the denominator counts the total number of terms in the numerator. The summations are performed over the rescaled discrete spatial coordinates. An example of calculating $\mathcal{D}$ is shown in the inset of Fig.~\ref{exponent}. We fix $L_1=15$, $s_1=10.67~\text{MHz}/\mathrm{\upmu s}$ and $L_2=23$ and vary $s_2$ to change $\mu$. For each $\mu$, we can determine $\mathcal{D}$ using the procedure described above. The results are shown in Fig.~\ref{exponent}. We fit the dataset $\{1/\mu, \mathcal{D}\}$ with a polynomial function and then determine the minimum of the fitting function. The minimum value of $1/\mu_\text{c}$ corresponds to the situation where two curves overlap most perfectly, yielding the desired critical exponent. In this case, the polynomial function is chosen as a cubic function because it adequately fits all data points without overfitting. To account for the error bar of $\mathcal{D}$, we employ a Monte Carlo method to simulate the uncertainties of $\mathcal{D}$ of each data point~(see \cite{supplementary}). By repeating the Monte Carlo sampling for $10^6$ times, we generate an ensemble of the fitted cubic functions, from which we calculate the error bar of the optimal $1/\mu_\text{c}$. In this instance, we find $\mu_\text{c}=0.50(5)$.

\begin{figure}
    \begin{center}  
    \includegraphics[width=0.86\columnwidth]{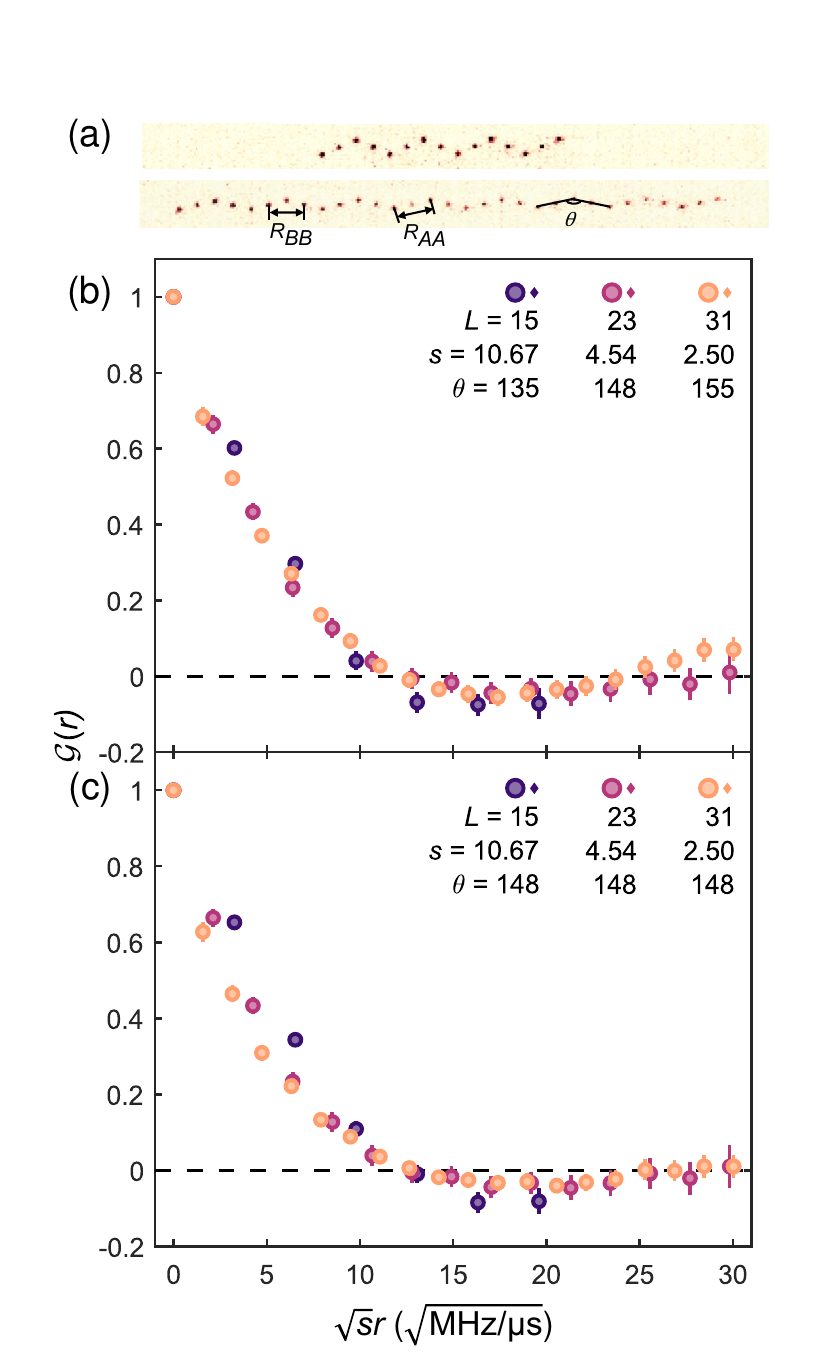}
    \caption{Scaling of the perturbation field in a zigzag geometry. (a) Images of a one-dimensional zigzag atom array with two different system sizes and zigzag angles. (b) The same correlation function $\mathcal{G}(r)$ plotted as a function of $\sqrt{s}r$ for the same set of system sizes and ramping speeds as Fig.~\ref{comparison}(a), except for the zigzag geometry with the zigzag angles varying with ramping speeds to maintain $h^{8/15}/\sqrt{s}$ constant. (\textbf{c}) The same plot as (\textbf{b}) except that $\theta=148^\circ$ is fixed.   } 
    \label{zigzag}
    \end{center}
\end{figure}

\textit{Effect of Symmetry-Breaking Field.} In this phase transition, the symmetry in question is the lattice translation symmetry, which can be explicitly broken by arranging atoms in a zigzag geometry, as illustrated in Fig.~\ref{zigzag}(a). In the zigzag configuration, atoms positioned at the corner, referred to as site $A$, differ from those located at the edge, referred to as site $B$, resulting in two distinct next-nearest distances. The distance between two $A$ sites, $R_{AA}$, remains $2a$, while the distance between two $B$ sites, $R_{BB}$, becomes $2a\sin(\theta/2)$, where $\theta$ is the zigzag angle. Consequently, this leads to two different next-nearest Rydberg interaction energies, with their difference $h$ proportional to $\frac{1}{2^6}\left(\frac{1}{\sin^6(\theta/2)}-1\right)$. Because of this energy difference, the two degenerate configurations depicted in Fig.~\ref{illustration}(d) are no longer degenerate. Thus, the energy difference $h$ is reminiscent of the longitudinal field in the transverse field Ising model. In the presence of a finite field $h$, there is no critical point, only a smooth crossover. For the Ising transition, it is known that $\nu_h=8/15$~\cite{yang1952spontaneous,zamolodchikov1989integrable,coldea2010quantum}.

Experimentally, we utilize the reconfigurable advantage of optical tweezers to arrange atoms into a zigzag geometry. The position accuracy of the static tweezers, and thus of the zigzag angle, is improved by adopting the zero-padding method~(see \cite{supplementary}). We repeat the KZ scaling measurement described previously in the zigzag geometry, with the results shown in Fig.~\ref{zigzag}. First, it is noteworthy to compare Fig.~\ref{zigzag}(c) with Fig.~\ref{comparison}(a). The three curves in both figures share the exact parameters of system sizes and ramping speeds. However, the data collapse in Fig.~\ref{zigzag}(c) is noticeably worse than that shown in Fig.~\ref{comparison}(a). This aligns with the fact that the presence of finite $h$ due to zigzag geometry breaks the quantum criticality from a different perspective, resulting in deviations from KZ scaling that cannot be restored even though the system size is already correctly scaled. Next, using the relationship between $h$ and $\theta$, we calculate that for these three system sizes and ramping speeds, the corresponding zigzag angles $\theta$ should be chosen as $135^\circ$, $148^\circ$ and $155^\circ$, respectively, in order to scale $h$ as $s^{\mu/\nu_h}$. Fig.~\ref{zigzag}(b) shows the results where $\theta$ varies following the change of $s$, considerably restoring the degree of data collapse. 

\let\oldaddcontentsline\addcontentsline
\renewcommand{\addcontentsline}[3]{}

\textit{Summary and Outlook.} In summary, we have experimentally demonstrated a novel protocol that allows us to achieve precise KZ scaling immune from finite-size effects and weak symmetry-breaking perturbations. This demonstration sharpens KZ scaling as a powerful tool for studying quantum criticality. While this work focuses on phase transitions driven by a single parameter, it can be straightforwardly generalized to phase diagrams controlled by multiple parameters. One of the most interesting phenomena of quantum criticality is the emergent exotic symmetry near some critical points. Examples include emergent Lorentz symmetry in the Bose-Hubbard model~\cite{greiner2002quantum,zhai2021ultracold} and emergent supersymmetry in the ladder of dual-species Rydberg atom arrays~\cite{supersymmetry}. Such critical points with emergent symmetry are usually fine-tuned isolated critical points in a multi-dimensional phase diagram. Tuning the ramping trajectory to cross the isolated point exactly is usually an experimental challenge. Generalizing our protocol to these situations can help overcome this challenge and offers the potential to measure the universal critical exponents in a more experimentally accessible way.

\begin{acknowledgments}
We thank Yingfei Gu and Xingyu Li for the discussions and Jilai Ye for technical support. This work is supported by National Natural Science Foundation of China under Grant No.~92165203 (W.C.), No.~92476110 (J.H.), No.~12488301(H.Z.) and No.~U23A6004 (H.Z.),  National Key Research and Development Program of China No.~2021YFA1400904 (W.C.), No.~2021YFA0718303 (J.H.), No.~2023YFA1406702 (W.C. and H.Z.), Tsinghua University Initiative Scientific Research Program and Dushi Program (C.L. J.H. H.Z. W.C.), the XPLORER Prize (H.Z.), China Postdoctoral Science Foundation under Grant No.~2024M751609 (H.W.) and Postdoctoral Fellowship Program of CPSF under Grant No.~GZC20231364 (H.W.).
\end{acknowledgments}

\textit{Data and code availability.} Data and code are available on Zenodo~\cite{zhang2025zenodo}. The DMRG and TEBD calculations are performed using the ITensor library~\cite{ITensor}.

\vspace{0.1in}

\textbf{Appendix}

\textit{Theoretical Protocol for Near-Critical Kibble-Zurek Scaling.} We first briefly review KZ scaling in an infinitely large system~\cite{zurek1985cosmological,zurek1996cosmological,laguna1997density,del2014universality}. The time scale $\tau$ associated with order parameter dynamics diverges at $g_\text{c}$ with a universal scaling form as $\tau=\tau_0/\delta g^{z\nu}$, as shown by the dashed lines in Fig.~\ref{illustration}(b). Here $\tau_0$ is a non-universal constant. We consider a linear ramp that $\delta g$ changes as $st$ with a constant ramping speed $s$. There is a finite duration $\Delta t$ where the parameter holds around $g$, and $\Delta t$ is given by $\delta g/(d \delta g/dt)=t$ for a linear ramp ~\cite{zurek1985cosmological,zurek1996cosmological,laguna1997density,del2014universality}. The key ingredient of KZ scaling is to consider the equation $\tau=\Delta t$, namely, $\tau_0/(st)^{z\nu}=t$, and its solution $t^*$ is the intersection between the red solid line and the grey dashed line in Fig.~\ref{illustration}(b), which sets a crucial time scale. When $t\ll t^*$, $\tau\gg \Delta t$ and the dynamics is frozen. When $t \gg t^*$, $\tau\ll \Delta t$ and the dynamics become adiabatic. The essential approximation to obtain the KZ scaling is that the order parameter formation dynamics mainly occur around $t^*$, and therefore, the formed order parameter has an average domain size given by the characteristic length scale $\xi$ at $t^*$ ~\cite{zurek1985cosmological,zurek1996cosmological,laguna1997density,del2014universality}. Since $\xi$ scales as $\sim 1/\delta g^{\nu}$, the order parameter has a typical domain size $\sim 1/(st^*)^{\nu}\sim 1/s^\mu$, where $\mu=\nu/(1+z\nu)$. The domain size sets the characteristic decay length scale of the two-point correlation function ~\cite{laguna1997density}, resulting in KZ scaling.  

In a finite-size system, although the ground state length scale $\xi$ no longer diverges, it still obeys a universal function form as follows~\cite{CardyBook},  
\begin{equation}
\xi=L \mathcal{F}(\delta g L^{1/\nu}), \label{xi2}
\end{equation}  
and $\mathcal{F}(x)$ is a universal function that is regular at $x=0$ and approaches $1/x^{\nu}$ for $x\gg 1$, as illustrated by the blue solid line in Fig.~\ref{illustration}(b).
Because of the time scale $\tau \sim \xi^z$ and $\delta g=st$, we have 
\begin{equation}
\tau=L^{z} [\mathcal{F}(st L^{1/\nu})]^z.
\end{equation}
Following the same reasoning of KZ scaling~\cite{zurek1985cosmological,zurek1996cosmological,laguna1997density,del2014universality}, we still consider the equation $\tau=\Delta t$ to determine $t^*$. We can compare the position where the red solid line ($\Delta t$) intersects with the grey dashed line ($\tau$ at infinite system size) and with the blue solid line ($\tau$ at finite system size) in Fig.~\ref{illustration}(b), it is easy to see that, for a given system size $L$, the difference becomes larger when the ramping speed is slower. Because of this finite size effect, $\xi$ will gradually be smaller than $1/s^\mu$ scaling when $s$ decreases.  

To eliminate this deviation due to the finite size effect, we reconsider the $\tau=\Delta t$ equation at finite system size, that is, 
\begin{equation}
L^{z} [\mathcal{F}(st L^{1/\nu})]^z=t. \label{frozen}
\end{equation}
Note that we can recast Eq.~(\ref{frozen}) as
\begin{equation}
(s^\mu L)^{z} [\mathcal{F}(\tilde{t} (s^\mu L)^{1/\nu})]^z=\tilde{t}, \label{frozen2}
\end{equation}
where we define $\tilde{t}=s^{\mu z}t$ and we have used the relation $z\mu+\mu/\nu=1$. Therefore, the solution $\tilde{t}^*$ for Eq.~(\ref{frozen2}) is determined by $s^\mu L$ only. Meanwhile, we note that the lengths scale $\xi$ at $\tilde{t}^*$ is given by 
\begin{equation}
\xi=\frac{1}{s^{\mu}}(s^\mu L)\mathcal{F}(\tilde{t}^*(s^\mu L)^{1/\nu}).
\end{equation}
Hence, once $s^\mu L$ is fixed, the factor $(s^\mu L)\mathcal{F}(\tilde{t}^*(s^\mu L)^{1/\nu})$ is a constant. Thus, the conclusion is that $\xi$ varies as $1/s^\mu$ only when $s^\mu L$ remains fixed.  This result is consistent with the dimension analysis arguments ~\cite{Sandvik2011,rossini2021coherent,de2023out}.
This conclusion does not rely on any prior knowledge of the functional form of $\mathcal{F}(x)$. 

In the presence of symmetry-breaking field $h$, $\tau$ becomes
$\tau=L^z[\mathcal{F}(\delta g L^{1/\nu},hL^{1/\nu_h})]^z$. However, since another argument $hL^{1/\nu_h}$ does not depend on time $t$, and therefore, it does not affect the discussion above, as long as $hL^{1/\nu_h}$ is fixed as a constant.

\bibliography{reference.bib}

\clearpage
\onecolumngrid
\appendix
\subsection*{\Large Supplementary Materials}

\let\addcontentsline\oldaddcontentsline
\renewcommand{\theequation}{S\arabic{equation}}
\renewcommand{\thefigure}{S\arabic{figure}}
\renewcommand{\thesection}{S\Roman{section}}
\renewcommand{\thesection}{\arabic{section}}
\renewcommand{\thesubsection}{\Alph{subsection}}
\renewcommand{\thesubsubsection}{\arabic{subsubsection}}

\setcounter{equation}{0}
\setcounter{figure}{0}
\setcounter{table}{0}
\setcounter{section}{0}

\tableofcontents
\setcounter{secnumdepth}{3}

\newpage

\section{DETAILS ON EXPERIMENTS}
\subsection{Experimental system}
\begin{figure*}
    \begin{center}  
    \includegraphics[width=0.8\columnwidth]{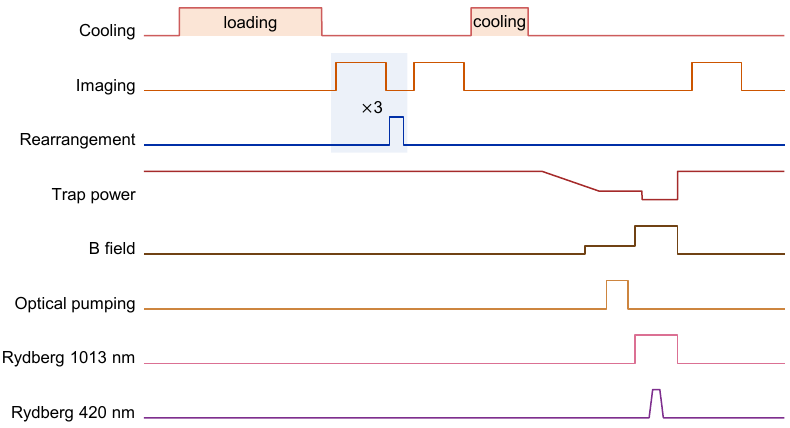}
    \caption{Experimental sequence. The experimental sequence begins with loading of single atoms into a set of static tweezers, followed by three cycles of rearrangements to arrange the atoms into the desired geometry. A fluorescence image is then taken to confirm that the 1D straight chain or the zigzag chain is defect-free. Then we perform adiabatic cooling by gradually lowering the tweezer depth. A magnetic field is switch on and defines the quantization axis. The atoms are then optically pumped into $\ket{g}$. After the state initialization, we further raise the magnetic field and switch off the tweezer. Then, we apply programmed Rydberg laser pulse and to drive the system across the phase transition point. Finally, the tweezers are turned back on and perform a final round of image for readout.
    } 
    \label{sequence}
    \end{center}
\end{figure*}

The experiments are enabled by a two-dimensional programmable quantum simulator based on a $^{87}$Rb atom array platform. 
The platform uses a set of static tweezers to load and hold the single atoms, and a set of mobile tweezers to rearrange the atoms.
We spatially modulate a 813-nm fiber laser beam using a spatial light modulator (SLM) (Holoeye, PLUTO-2.1-NIR-118) to create an arbitrary static tweezer array. The trap depth is set to $\sim 1.2~\mK$.
The mobile tweezers are generated using a pair of acousto-optic deflectors (AODs) (AA Opto Electronic DTSXY-400), which deflect the the beam from a 852-nm laser source (UniQuanta). 
Both sets of tweezers are focused by a microscope objective with numerical aperture ${\rm NA} = 0.5$ (Mitutoyo G Plan Apo 50X). 
A second objective collects fluorescence from atoms and monitors the tweezer array.
After the rearrangements, the atoms are cooled and initialized in the ground state hyperfine level. 
Then we couple the atoms to the Rydberg state and ramp the system across the phase transition, followed by the detection of the Rydberg state occupation.
The overall experimental sequence is shown in Fig.~\ref{sequence}, with details described below.

\subsubsection{Static tweezer generation for the straight \& zigzag chain}
\begin{figure}
    \begin{center}  
    \includegraphics[width=0.5\columnwidth]{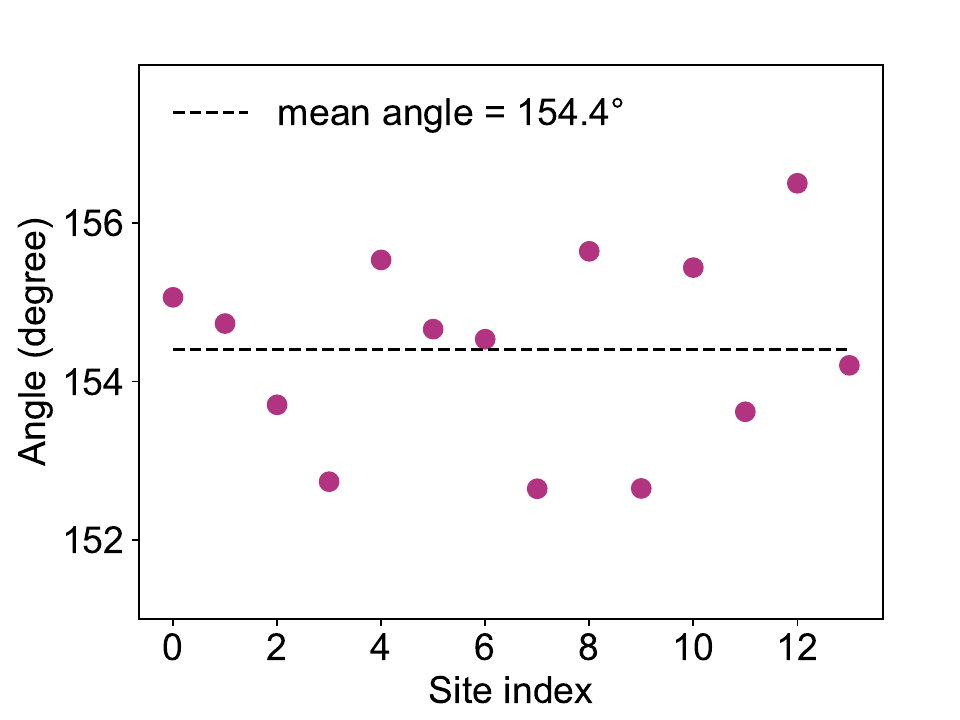}
    \caption{Measurement of zigzag angles. As an example, we show the measured zigzag angle for a target angle of $155\degr$. Each data point indicates a measured angle fitted from the image on the tweezer monitoring camera. The mean angle is $154.4\degr$, and the standard deviation is $1.2\degr$.
    } 
    \label{zigzag_angle}
    \end{center}
\end{figure}

The hologram imprinted on the SLM is computer-generated using the phase-fixed weighted Gerchberg-Saxton (WGS) algorithm to produce an arbitrary configuration of tweezers \cite{kim2019gerchberg}. 
We measure the intensity of each tweezer and iteratively optimize the hologram to achieve $<5\%$ intensity inhomogeneity across all tweezer sites both for the 1D straight chain and for the zigzag geometry~\cite{kim2019large}.
For the zigzag geometry, a critical challenge arises from the uncertainty of tweezer positions, particularly when dealing with large zigzag angles. 
The minimum resolvable tweezer position, limited by the SLM resolution ($1920\times 1080$), is insufficient for our target zigzag configuration.
To enhance the precision of tweezer position, we apply the zero-padding with a padding factor of 10 during phase pattern generation and intensity homogeneity feedback. 
The performance is evaluated by fitting the tweezer positions obtained from the tweezer monitor camera to extract the zigzag angles (Fig.~\ref{zigzag_angle}).
We find that the zigzag angle deviation converges to $\sim 1.2\degr$.\\

\subsubsection{Rearrangement protocol}
\begin{figure}
    \begin{center}  
    \includegraphics[width=0.4\columnwidth]{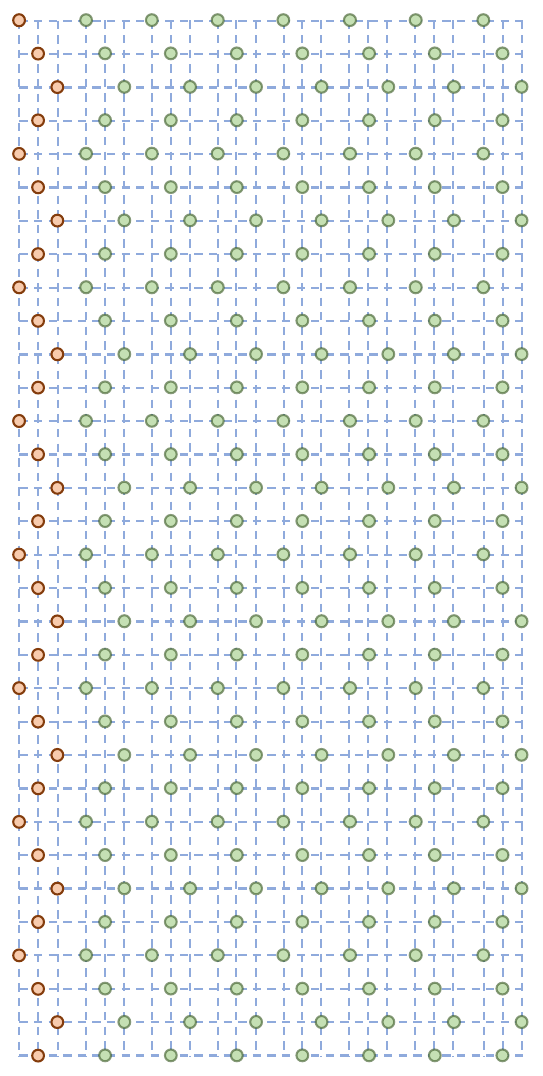}
    \caption{Rearrangement for the zigzag chain. The target sites are positioned in the leftmost zigzag chain, while other 7 identical copies serve as reservoirs. They are pixelated as a subset in a $32 \times 24$ lattice structure. The horizontal and vertical dashed blue lines denote the corresponding lattice coordinates as encoded in the FPGA implementation.
    } 
    \label{zigzag_config}
    \end{center}
\end{figure}

While in most part of the experiment we use the labscript suite \cite{Starkey_2013} to control the experiment, the rearrangement protocol is implemented by an FPGA-based hardware, as detailed in Ref.~\cite{wang2023accelerating}.
For 1D straight chains, the atoms are initially loaded into a $32 \times 8$ reservoir tweezer array, with the target sites positioned in the leftmost column. 
Following the fluorescence imaging, we firstly apply a row-by-row rearrangement to move one atom to the leftmost column and other atoms to the rightmost area. 
Then for each column we rearrange extra atoms to the rows with empty target sites to prepare for the next rearrangement operation. 
We repeat the rearrangement for three times consecutively in each cycle to improve the rearrangement success rate.
In the last rearrangement trial, we remove all the redundant atoms.
For the zigzag geometry, we arrange the static tweezer array as 8 parallel copies of the target zigzag chain, with the target sites positioned in the leftmost chain. The entire static array is pixelated as a subset in a $32 \times 24$ lattice recognized by the FPGA, as shown in Fig.~\ref{zigzag_config}. When the zigzag angle is large, the distance between two columns is very small, resulting in a high probability of atom loss during the vertical movement. To mitigate this, we switch off column movement and only perform row movement. In the row movement, instead of moving all extra atoms to the rightmost area, we leave them at their original positions, except for the last trial, in case of needing them for the next rearrangement. Three trials are also performed in each cycle.
For all data presented in the main text, we post-select shots with defect-free atom arrays.\\ 

\subsubsection{Atom temperature}
After the rearrangement, we switch on the cooling lasers again to further lower the atom temperature. Then, we implement the adiabatic cooling by ramping down the tweezer trap power to $\sim 1/6$ of that in the loading stage. 
By performing a release-and-recapture experiment, the atom temperature is measured to be $\sim 15.5~\uK$.\\

\subsubsection{State preparation}
We initialize the atomic state by optical pumping into the ground state hyperfine level $\ket{g}=\ket{5S_{1/2},F=2, m_F=+2}$ with a $\sigma+$ pumping beam that propagates parallel to the quantization axis, which is set by a 1.5-Gauss magnetic field.
The pumping beam contains both repumper light on the $F=1 \rightarrow F^{\prime}=2$ transition and optical pumping light on the $F=2 \rightarrow F^{\prime}=2$ transition. The preparation fidelity is estimated to be $>98\%$, including the atom loss $\sim 0.9(3)\%$ and the pumping fidelity $>99\%$.\\

\subsubsection{Rydberg state coupling}
\begin{figure}
    \begin{center}  
    \includegraphics[width=0.5\columnwidth]{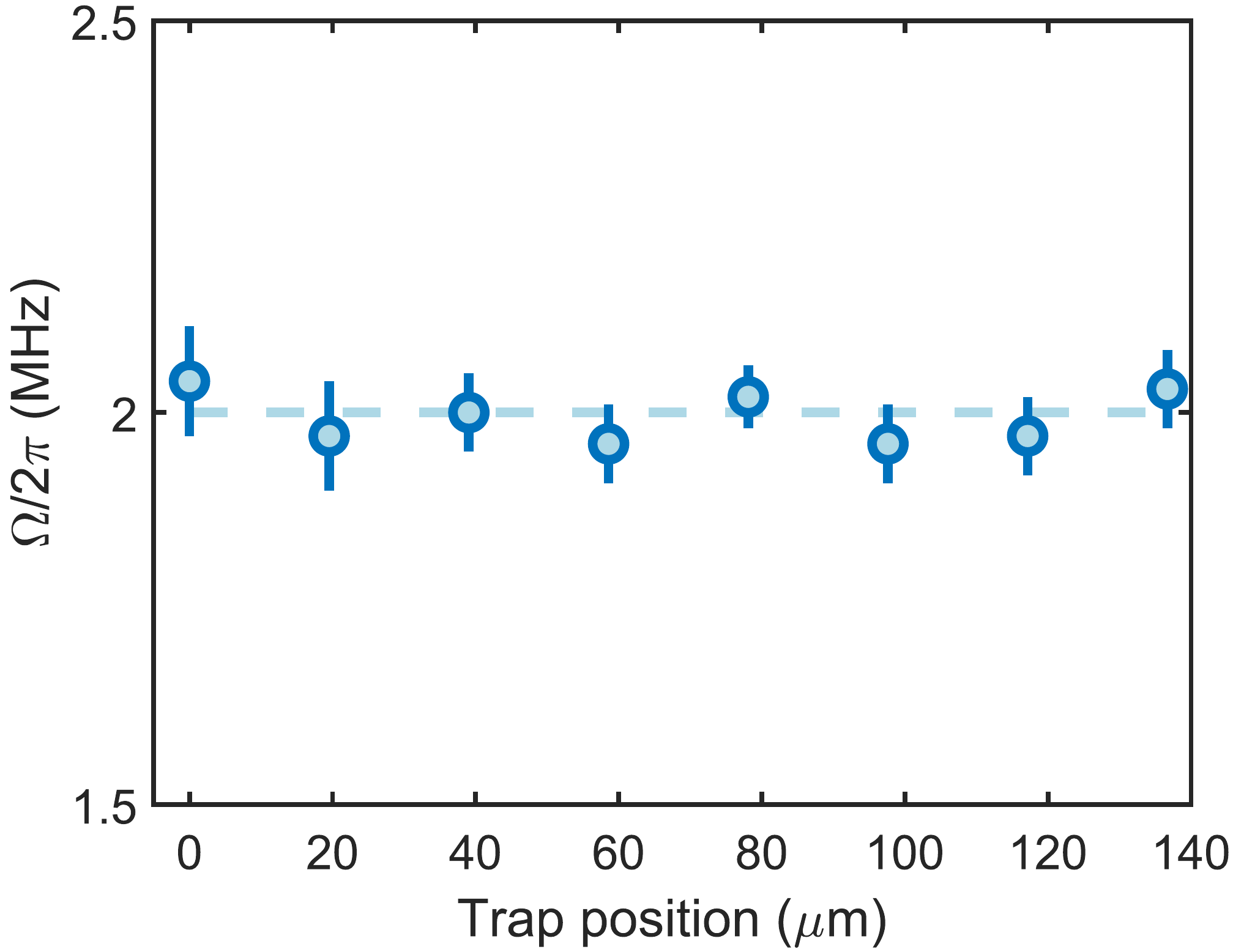}
    \caption{Inhomogeneity of Rydberg coupling Rabi frequencies accross the atom chain. Rabi frequency as a function of trap position along the direction of the chain used in the experiments in this work. Each data point represent the Rabi frequency fitted from a Rabi oscillation measurement. The data shows a standard deviation of $2\pi \times 0.063~\MHz$, or $3\%$, of the Rabi frequencies over a range of $150~\um$.
    } 
    \label{Rabi}
    \end{center}
\end{figure}

We couple the ground state $\ket{g}$ to the Rydberg state $\ket{r}=\ket{68S_{1/2}, m_J=+\frac{1}{2}}$ via a two-photon transition through the intermediate state $\ket{e}=\ket{6P_{3/2}, F=3, m_F=+3}$ with a single-photon detuning of $\sim 2\pi \times 700~\MHz$. 
The 420~nm laser (Precilaser, fiber laser) drives the $\ket{g} \rightarrow \ket{e}$ transition, and the 1013~nm laser (cat-eye diode laser, amplified by a fiber amplifier, both from Precilaser) drives the $\ket{e} \rightarrow \ket{r}$ transition. 
With incident powers of $8~\mW$ and $1~\W$, respectively, the two-photon Rabi frequency is set to $2\pi\times2~\MHz$. 
The lasers are locked to ultra-low expansion (ULE) cavities (Stable Laser Systems) with finesse $\mathcal{F}_{420\text{-}\mathrm{nm}}=3000$ and $\mathcal{F}_{1013\text{-}\mathrm{nm}}=10000$. 
The frequency and amplitude of the 420-nm laser are further modulated by a pair of acousto-optic modulators (AOM, AA OPTO-ELECTRONIC, MQ240-A0,2-UV) with a short rise time of 20~ns. 
One of the AOMs is driven by an arbitrary waveform generator (AWG, M4i.6631-x8) to enable fast switching and frequency modulation. 
We use photodetectors to monitor the Rydberg laser power and correct for the pulse profile by calibrating the AWG-generated control signal.
Both laser beams are shaped into a stripe-type profile with a waist of $14~\um \times 500~\um$ for the 420-nm laser and $25~\um \times 125~\um$ for the 1013-nm laser by using cylindrical lenses.
The atom chain is aligned along the laser propagation direction to minimize the inhomogeneity of the Rabi frequency. 
As shown in Fig.~\ref{Rabi}, the standard deviation of Rabi frequencies is found to be $2\pi \times 0.063~\MHz$, or $3\%$, across the chain. 
To avoid beam drift resulting in a change of the Rabi frequency during the long data taking time, we monitor the pointing with a CMOS camera and feedback the beam pointing with a piezo mirror to maintain a consistent Rabi frequency.
During the Rydberg excitation, we apply a magnetic field of 4~Gauss along the chain to mitigate the effect of the stray magnetic field on the Rydberg excitation.\\

\subsubsection{Rydberg interaction measurement}
\begin{figure}
    \begin{center}  
    \includegraphics[width=0.5\columnwidth]{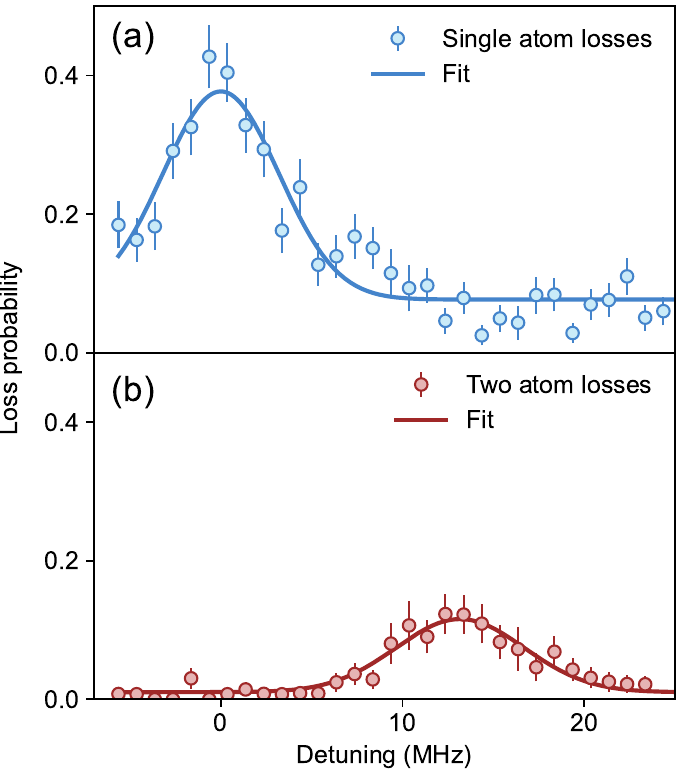}
    \caption{Spectroscopic measurement of Rydberg interactions and inter-atomic distances.
    (\textbf{a}) Rydberg excitation spectrum for single-atom loss. The peak corresponds to the two-photon coupling from $\ket{g,g}$ to $\ket{W}=(\ket{g,r}+\ket{r,g})/\sqrt{2}$. (\textbf{b}) Rydberg excitation spectrum for two-atom loss. The peak corresponds to the four-photon coupling from $\ket{g,g}$ to  $\ket{r,r}$, which is shifted by the interaction energy $V_{\rm NN}/2$ from the single-atom loss peak. In both figures, solid lines are fits into Gaussians with offset.} 
    \label{rydberg_interaction}
    \end{center}
\end{figure}

We measure the interaction between nearest-neighbor atoms by comparing the resonant frequency of the two-atom loss peaks in the Rydberg excitation spectrum (bottom in Fig.~\ref{rydberg_interaction}) to the single-atom loss peak (top in Fig.~\ref{rydberg_interaction}). Due to the van der Waals interaction between Rydberg atoms, the two-atom loss peak is shifted from the single-atom loss peak by the interaction energy $V_{\rm NN}/2$. The nearest-neighbor interaction is determined to be $V_{\rm NN}=2\pi \times 26.4(4)~\MHz$, which gives the interatomic distance $5.34(2)~\um$.\\

\subsection{Data analysis of $\mathcal{G}(r)$}
We measure the final state of every atom by identifying absence of atom in the fluorescence imaging as the atom being in the Rydberg state. In theory, the Rydberg blockade effect prevents two Rydberg atoms from being excited at neighboring sites. In practice, however, we observe double excitations from time to time due to the finite probability of atom loss during the experiments. Therefore, we discard the shots that violate the Rydberg blockade rule. The correlation function is given by
\begin{equation}
    G(r) = \frac{1}{\mathcal{N}(r)}\sum_i (\langle \hat{n}_i \hat{n}_{i+r}\rangle -\langle \hat{n}_i\rangle \langle \hat{n}_{i+r}\rangle),
\end{equation}
where $\langle \cdots \rangle$ denotes an ensemble average, and $\mathcal{N}(r)$ is the number of site pairs separated by distance $r$. To reduce boundary effects, we remove four sites from each end of the chain, thus adjusting $\mathcal{N}(r)$ to $L-r-8$. The correlation function is then rescaled by $G(0)$ to obtain $\mathcal{G}(r) = (-1)^r G(r)/G(0)$.

We quantify the statistical uncertainty in the correlation measurements by using the bootstrapping method. Specifically, we generate over a thousand bootstrap samples by randomly resampling the original data with replacement. From these resampled datasets, we calculate $\mathcal{G}(r)$ for each bootstrap sample, and then obtain both the mean value and standard deviation by averaging over all bootstrap samples. \\

\subsection{Data analysis of $\mathcal{D}$}
As described in the main text, we use the distance function $\mathcal{D}$ to quantify how well two correlation functions, after appropriate rescaling, collapse into a single curve. It is defined as \cite{bhattacharjee2001measure}
\begin{equation}
\mathcal{D} = \frac{\sum_i |\mathcal{G}_1(\tilde{r}_i) - \mathcal{G}_2^{\text{lin}}(\tilde{r}_i)| + \sum_j |\mathcal{G}_2(\tilde{r}_j) - \mathcal{G}_1^{\text{lin}}(\tilde{r}_j)|}{N_{1} + N_{2}}.
\end{equation}
The summations are performed over the set of rescaled spatial coordinates.
For Fig.~3 in the main text, we pick the data points of $\mathcal{G}(r)$ with $r\leq5$ for $L=15$ to reduce error contributed by statistical error.
This symmetric measure of deviation provides a robust way to assess the quality of the data collapse for a given scaling exponent $\mu$.

To estimate the uncertainty of $\mathcal{D}$, we employ a Monte Carlo approach by assuming that each data point follows an independent Gaussian distribution according to the mean value and uncertainty of the correlation function. We generate $10^6$ sample datasets and calculate the corresponding $\mathcal{D}$ for each sample. The standard error of $\mathcal{D}$ is then obtained by taking the standard deviation of the $10^6$ results.\\

\section{DETAILS ON NUMERICAL SIMULATIONS}
\subsection{Finite size scaling of the ground state}
To verify the critical exponents at the critical point, we perform standard finite-size scaling for the ground state using a finite-system density-matrix renormalization group (DMRG) algorithm. After extracting the two-point correlation functions $\mathcal{G}(r)$ numerically, we compute the correlation length by \cite{landau2021guide}
\begin{equation}
  \xi = \frac{1}{2 \sin\bigl(2\pi / L\bigr)} \,
  \sqrt{\frac{\chi(0)}{\chi\bigl(4\pi / L\bigr)} - 1},
\end{equation}
where
\begin{equation}
  \chi(k) \;=\; \sum_{r \,\in\, \text{even sites}} 
  \mathcal{G}(r)\, e^{i k r}.
\end{equation}

\begin{figure}[t]
  \centering
  \includegraphics[width=0.5\columnwidth]{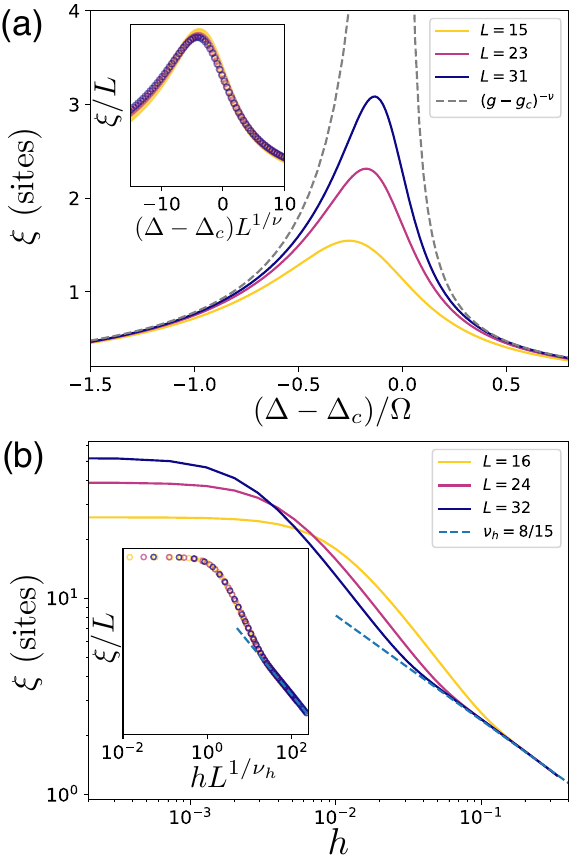}
  \caption{Finite-size correlation length of the ground state.
  (a) Correlation length $\xi$ vs detuning around the $Z_2$ critical point. Inset: data collapse with $\nu=1$.
  (b) Correlation length $\xi$ vs the effective longitudinal field $h$. Inset: data collapse with $\nu_h=8/15$.}
  \label{fig:correlation_length_equilibrium}
\end{figure}

Fig.~\ref{fig:correlation_length_equilibrium}(a) shows $\xi$ vs the detuning (in units of $\Omega$). For different system sizes, data far away from the critical point collapse onto a single curve, consistent with a power-law $\xi \sim 1/(g - g_c)^\nu$, with $g = \Delta/\Omega$ and $\nu = 1$. This agrees with the $Z_2$ critical exponent in the thermodynamic limit. Near the critical point, finite-size effects become more pronounced: smaller systems deviate more from the infinite-size prediction. Nevertheless, the standard finite-size scaling form 
$\xi/L = \mathcal{F}_\perp\bigl((g - g_c)\,L^\nu\bigr)$
holds, as demonstrated by the inset of Fig.~\ref{fig:correlation_length_equilibrium}(a).

\begin{figure}[t]
  \centering
  \includegraphics[width=0.5\columnwidth]{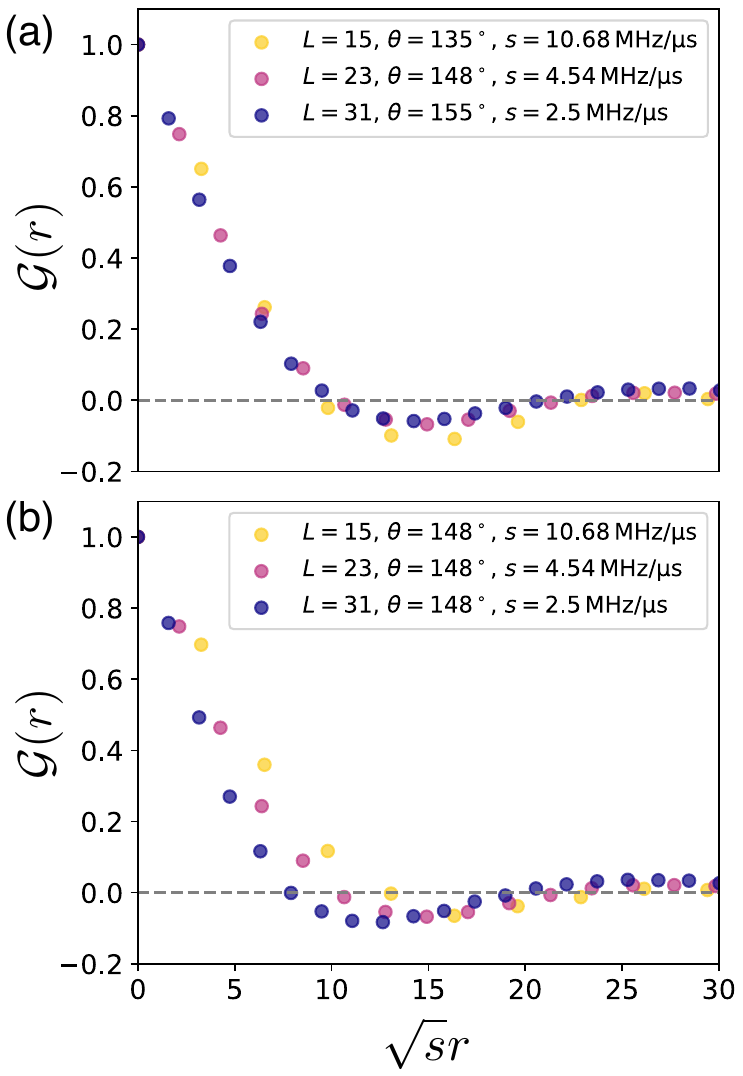}
  \caption{Numerical simulation of KZ dynamics in a zigzag geometry.
  We set $R_b/a = 1.4$. Panel (a) shows the scaling behavior of $\mathcal{G}(r)$ for a finite zigzag angle at various system sizes, while panel (b) illustrates the corresponding uniform longitudinal-field effect for comparison.}
  \label{fig:zigzag_KZ}
\end{figure}

We also examine the ground state in the presence of a zigzag geometry (Fig.~4(a) in the main text) and attempt to identify its corresponding critical exponent. This zigzag geometry induces both a longitudinal field and a transverse field \cite{wang2025lattice}. Specifically, it induces a (dimensionless) longitudinal field
$h(\theta) = \left(\frac{R_b}{2a}\right)^6 \left(\frac{1}{\sin^6(\theta/2)} - 1\right)$,
which arises from the difference between $R_{AA}$ and $R_{BB}$. Theory predicts that the correlation length decays with a scaling of $1/h^{\nu_h}$, where $\nu_h = 8/15$ in the thermodynamic limit. However, due to the presence of the transverse field effect, the original $g_c$ is shifted for any angle $\theta \neq 180^\circ$, and we define it as $g_c(\theta)$. Our goal is to determine the correlation length at $g_c(\theta)$ and verify how its power-law decay against $h(\theta)$, with $\nu_h = 8/15$, holds in the zigzag geometry.

To determine $g_c(\theta)$, we use a \emph{symmetric} zigzag arrangement where the next-nearest-neighbor distances remain uniform:
\begin{tikzpicture}[scale=0.5, line cap=round, line join=round, baseline=-2pt]
    \coordinate (A0) at (0,0);
    \coordinate (A1) at (0.5,0.5);
    \coordinate (A2) at (1,0);
    \coordinate (A3) at (1.5,0.5);
    \coordinate (A4) at (2,0);
    \draw[thick] (A0) -- (A1) -- (A2) -- (A3) -- (A4);
    \foreach \point in {A0, A1, A2, A3, A4} {
      \draw[fill=black] (\point) circle (3pt);
    }
\end{tikzpicture}
. This configuration only includes the transverse effect, which is twice as strong as in the zigzag geometry with an AB-sublattice
\begin{tikzpicture}[scale=0.5, line cap=round, line join=round, baseline=-2pt]
    \coordinate (A0) at (0,0);
    \coordinate (A1) at (0.25,0.25);
    \coordinate (A2) at (0.5,0.5);
    \coordinate (A3) at (0.75,0.25);
    \coordinate (A4) at (1,0);
    \coordinate (A5) at (1.25,0.25);
    \coordinate (A6) at (1.5,0.5);
    \coordinate (A7) at (1.75,0.25);
    \coordinate (A8) at (2,0);
    \draw[thick] (A0) -- (A1)-- (A2) -- (A3) -- (A4) -- (A5)-- (A6)-- (A7)-- (A8);
    \foreach \point in {A0, A1, A2, A3, A4, A5,A6,A7,A8} {
      \draw[fill=black] (\point) circle (3pt);
    }
\end{tikzpicture}
, if these two types of zigzags share the same nearest-neighbor spacing $a$ and kink angle $\theta$. Therefore, we have $g_c(\theta)=g_c(180\degr) + \left(g_c^\text{symm}(\theta)-g_c(180\degr)\right)/2$, where $g_c^\text{symm}(\theta)$ can be determined by the Binder ratio in the symmetric zigzag case. We now go back to calculate the ground state wavefunction at $g_c(\theta)$, and find that $\xi(\theta)$ vs $h(\theta)$ exhibits the expected scaling behavior (Fig.~\ref{fig:correlation_length_equilibrium}(b)). The inset confirms the corresponding finite-size scaling 
$\xi/L = \mathcal{F}_\|(hL^{1/\nu_h}).$
We argue that this transverse field effect does not affect our KZ protocol for detecting $\nu_h$, as it merely shifts the crossover center $g_c(\theta)$, which will regardless be crossed when the detuning is ramped in a dynamical protocol.

Taking into account both finite-size deviations and the zigzag-induced fields, we conclude that the correlation length obeys a two-parameter scaling of the form
\begin{equation}
    \xi = L\mathcal{F}\left(\left(g - g_c(\theta)\right)L^{1/\nu},h(\theta)L^{1/\nu_h}\right).
\end{equation}

\subsection{Numerical simulation of KZ dynamics}
We employ the time-evolving block decimation (TEBD) algorithm to simulate KZ dynamics. We include interactions up to $\le 6$ sites among the atoms and use a Trotter decomposition for the unitary evolution, following Ref.~\cite{keesling2019quantum}. The system is driven by the same pulse shape used in the experiment, with time steps of $0.001\,\upmu\mathrm{s}$ for switching the Rabi coupling on/off and $0.0002\,\upmu\mathrm{s}$ for ramping the detuning. We set an SVD truncation cutoff of $10^{-11}$. Using this approach, we simulate the correlation function $\mathcal{G}(r)$ (defined in the main text) for both a 1D straight chain ($\theta = 180^\circ$, as in 
Fig.~2 in the main text) and a finite zigzag angle, as shown in Fig.~\ref{fig:zigzag_KZ}.

\end{document}